# Beat Frequency Modulation of T Tauri Accretion Rates

Kester W. Smith, Ian A. Bonnell & Geraint F. Lewis
*Institute of Astronomy, Madingley Road, Cambridge CB3 0HA.*

**ABSTRACT**

A general model of magnetically controlled accretion onto T Tauri stars is presented. In this model the magnetic field is oriented arbitrarily in relation to the star's rotation axis. The resultant interplay between the magnetic field and accretion disc causes a variable accretion rate. The dominant timescale of this variability is the beat frequency between the stellar rotation frequency and the orbital frequency at the magnetosphere boundary. This model is analogous to that developed to explain quasi periodic oscillations in low-mass X-ray binaries.

## 1 INTRODUCTION

The current standard model of T Tauri stars is that they accrete material via an accretion disc. This disc is truncated a few stellar radii from the surface by a stellar magnetic field, which then channels the disc material onto the star's surface. This model was originally developed to provide the spin-down mechanism necessary to explain the modest rotation rates observed in T Tauri systems (Königl 1991). Periodic continuum modulation observed in a number of T Tauri systems has been interpreted as rotational modulation due to the presence of photospheric spots (Bouvier et al 1993; Bouvier et al 1994). These spots may be cool magnetic starspots, or hot regions associated with localised accretion shocks. Calvet & Hartmann (1992) have shown that the structure of several spectral lines can be explained in terms of the magnetically funnelled accretion model. The presence of Inverse P Cygni profiles in YY Orionis type stars, a subset of T Tauri stars, is a sign of infalling material which is associated with the magnetically-channeled accretion columns. Edwards et al (1994) found a high incidence of such profiles amongst a sample of T Tauri stars chosen to span a range of accretion rates.

Photometric studies by various authors have revealed periodic variability in the light curves of many T Tauri stars (Bouvier et al 1986; Bouvier & Bertout 1989; Vrba et al 1989; Vrba et al 1986). This is commonly interpreted as rotational modulation caused by the presence of surface spots on the photosphere. It has been shown that the periods of the so-called Classical T Tauri stars (CTTS), which show strong accretion signatures, tend to be significantly longer than those of Weak line T Tauris (WTTS), which have weaker accretion signatures (Attridge & Herbst 1992; Edwards et al 1993; Bouvier et al 1993). The obvious conclusion, that CTTS rotate slower than WTTS, is surprising, since accretion of high specific angular momentum material is expected to spin the star up. It has been suggested, (Clarke et al 1994), that varying magnetic field can lead to the accretion process being switched on and off, the magnetosphere boundary remaining close to the corotation radius, but moving back and forth across it as the magnetic field strength fluctuates. A T Tauri star could spin down due to magnetic braking during its accretion phase, whilst appearing sometimes as a WTTS, and sometimes as a CTTS. If this picture is correct, there would be no reason to expect a significantly different distribution of rotation periods between the two classes.

Previous studies of magnetically channeled accretion have limited themselves to the simple case where the magnetic axis is perfectly aligned with the rotation axis. At the same time, attempts to explain CTTS periodicity by invoking magnetically channeled accretion shocks imply a misaligned field. In this paper we generalise the model to include the effects of misalignment of magnetic and rotation axes, and show that, in the case of such a misalignment, the longer periods in CTTS light curves do not necessarily indicate longer rotation periods for the stars.

In Section 2, we outline the basic model, and discuss various immediate implications. In section 3 we review what is currently known of T Tauri periodicities and rotation. In Section 4, we discuss various consequences of beat frequency modulated accretion which could be tested observationally. Section 5 summarizes our conclusions.

## 2 THE BEAT FREQUENCY MODEL

### 2.1 Outline

Unless the magnetic and rotation axes of a star are perfectly aligned, the strength of the interaction between the material at the inner edge of the disc and the magnetic field will vary with azimuth. A perfectly dipolar field will produce two field strength maxima, and two intermediate field strength minima. A particular part of the inner disc will encounter the same maximum or minimum field strength region repeatedly, with a frequency;

$$\omega_{beat} = |\omega_m - \omega_*|$$

the beat frequency between the magnetospheric orbital frequency $\omega_m$ and the stellar rotation frequency.

Any inhomogeneity in the disc material will lead to a variable accretion rate onto the star. Non-axisymmetric inhomogeneity is anticipated for a number of reasons. Non-axisymmetric structure can be excited by the presence of a companion star outside the disc or a protoplanet forming within the disc. Magnetohydrodynamic instabilities can lead to the formation of clumps in the disc (Stella & Rosner 1984). Observational evidence for clumpy discs has also been found (Graham 1992).

The accretion rate will reach a maximum when a high density region of the inner disc coincides with a field strength maximum. If the disc is significantly denser at some azimuths, a periodic variation will be seen in the light curve. The frequency of this variation will be some harmonic of the beat frequency. A perfect dipole, and a disc with one dense region at its inner edge, will produce variability with frequency $2\omega_{beat}$. If the field is roughly dipolar, but with one pole stronger than the other, the dominant frequency observable in the light curve will be simply $\omega_{beat}$.

This model was developed by Alpar & Shaham (1985) to explain the quasi-periodic oscillations sometimes seen amongst Low mass X-ray binaries when in their so-called horizontal branch phase.

### 2.2 Consequences of modulated accretion

Applying this model to T Tauri stars offers an alternative explanation of the photometric periods observed, and has been briefly mentioned in terms of T Tauri stars (Bouvier et al 1994). Assuming the accretion and photospheric luminosities to be equal at V, a 20 percent increase in the accretion rate would be required to produce a variation of 0.1 magnitudes in the light curve. A clump capable of producing such an increase would be substantial and could be expected to last for a number of orbits. The accretion luminosity will form a greater fraction of the total luminosity at shorter wavelengths, so the variations would be expected to increase in amplitude at bluer wavelengths, and decrease in amplitude at redder wavelengths.

The actual beat frequency depends upon the radius of the magnetosphere, which determines $\omega_m$, and the stellar rotation frequency. The latter is expected to change only slowly, on a timescale of approximately $10^5$ years, (Königl 1991; Armitage 1995), and so any sudden change in the observed photometric period would have to arise from a change in the orbital frequency at the magnetosphere. This could occur because of a change in the stellar magnetic field strength, or because of a change in the accretion rate.

The transition from Keplerian disc to corotating magnetosphere would not occur at a sharp boundary, although the transition zone is expected to be narrow compared to the radius of the magnetosphere. For our illustration we follow various other authors (Königl 1991; Clarke et al 1994) and take as our magnetosphere boundary;

$$R_m = \left(\frac{B_*^2 R_*^6}{(GM_*)^{\frac{1}{2}} \dot{M}}\right)^{\frac{2}{7}}$$

where the magnetic torques balance the accretion torques. Material in the disc orbits with Keplerian frequency;

$$\omega_{disc} = \left(\frac{(GM_*)}{R^3}\right)^{\frac{1}{2}}$$

Combining these two equations we can obtain an expression for the Keplerian orbital frequency at the magnetosphere boundary;

$$\omega_m = \left(\frac{\dot{M}(GM_*)^{\frac{5}{3}}}{B_*^2 R_*^6}\right)^{\frac{3}{7}}$$

The critical parameters are the accretion rate and the stellar magnetic field strength. Figs 1 and 2 show the dependence of the beat period on each of these parameters for four different stellar rotation periods. Only values of the beat period for $R_m < R_{co}$ are graphed, as no accretion is possible if $R_m > R_{co}$. The stellar mass in each case is taken to be $1M_\odot$ and the stellar radius is $2R_\odot$, typical values for T Tauri stars.

The beat period depends on the distance between the corotation radius and the magnetosphere boundary. In principle the innermost edge of the disc could extend almost down to the star's surface. In this case, the beat period might be of order a few hours (the Keplerian orbital period near the stellar equator).

In the steady state case Ghosh & Lamb (1979 a,b) have argued that the system should evolve to its equilibrium state, with the magnetospheric radius lying just inside corotation. In a non-steady state, the magnetospheric radius could vary due to variation in either the magnetic field or the accretion rate. In the former case, Clarke et al (1994) and Armitage (1995) have argued that the magnetic field can vary on timescales of 10-20 years, but that the mean magnetospheric radius remains near corotation. Evidence for variable accretion rates gives timescales from $10^3$ years down to a few days (Hartmann et al 1993).

### 2.3 Accretion shock geometry

Inverse P Cygni (IPC) profiles are a characteristic of the subset of T Tauri stars called YY Orionis stars. These profiles are evidence of infalling material. The infall velocity is often several hundred kms$^{-1}$, and these features are taken to be the signatures of the columns of infalling disc material.

A recent study by Edwards et al (1994) revealed IPC structure in at least one emission line for 13 out of a sample of 15 CTTS. The velocity of the redshifted absorption ranged from 200-300 kms$^{-1}$. This high incidence of IPC structure indicates that high velocity infalling material is visible from most lines of sight. This mitigates against the simplest dipole accretion picture in which material is accreted exclusively onto a localized region near the magnetic pole. Instead, the field close to the star may be dominated by short-range, high multipole modes. The accreting material may then reach the stellar surface at more than just two localized regions. This would imply that the photometric periods seen in CTTS are not as likely due to rotational modulation of localised spots.

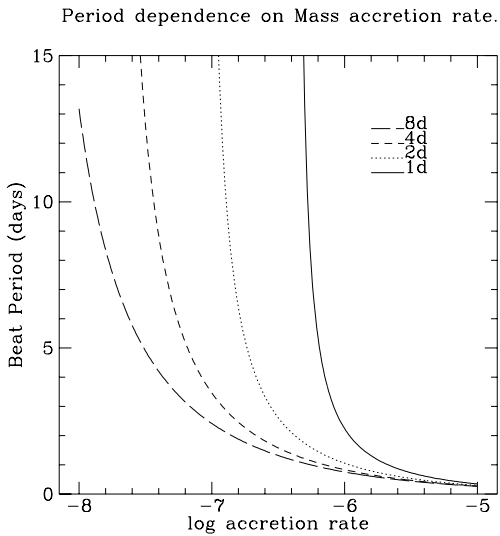

**Figure 1.** The dependence of the beat period on the accretion rate for 4 values of the stellar rotation period. The magnetic field is taken to be 750 Gauss.

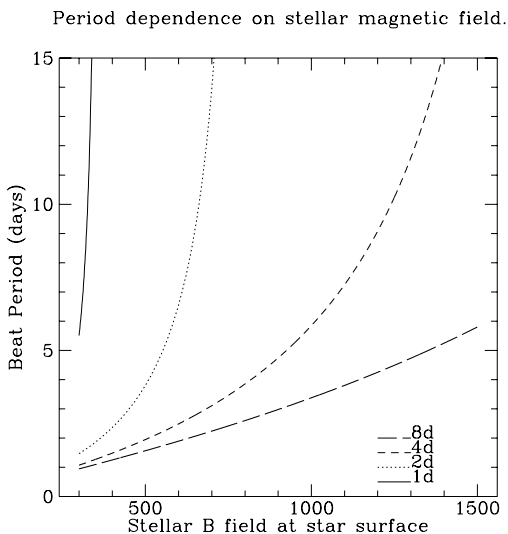

**Figure 2.** The dependence of the beat period on the stellar magnetic field for 4 values of the stellar rotation period. An accretion rate of $10^{-7} M_\odot \, \mathrm{yr}^{-1}$ is assumed.

## 3 PHOTOMETRIC PERIODS

There are a number of immediate consequences of the beat frequency model for the interpretation of photometric periods. The observed periods are then only indirectly related to the underlying rotation period. Firstly, it can be seen from section 2.1 above that the photometric period will change if either the accretion rate through the disk or the stellar magnetic field strength change.

There are a number of systems for which different photometric periods have been found at different times. RY Tau has been found to display periods of 7.25 days, 5.6 days (twice) and 24 days at various times (Chugainov et al 1991; Herbst et al 1987; Bouvier et al 1993). The 5.6 day period requires the radius measurement to be incorrect by a factor of over 3 to be consistent with the $v \sin i$ of 52.2 $\mathrm{kms}^{-1}$. Longer periods are in even worse agreement. BP Tau has had periods of 7.6 days and 6.1 days found in its light curve (Vrba et al 1986; Simon et al 1990). The latter also saw the periodicity change from one period to the other during an observing run. SY Cha has measured periods of 6.1 days and 7.6 days (Schaefer 1983; Kappelmann & Mauder 1981). These period changes have been seen exclusively in actively accreting CTTS. It is not possible to account for these period changes, if real, in terms of the rotational modulation picture, as the rotation period of the stars could not change on these timescales. However, it should be noted that in the case of CTTS, the likelihood of spurious period detections is greater than in WTTS, since CTTS display more random variability than WTTS in general.

Any significant change in the accretion rate would be observable as an increase in accretion luminosity, both from the disk and from any accretion shock. A modest change in the magnetic field strength might be expected to cause a short-lived increase or decrease in the accretion rate onto the star as the magnetospheric radius changes. If the magnetosphere moves out, extra material will be accreted as the disk is cleared out to a greater radius. If the magnetosphere moves in, accretion to the star will drop until the disk material can move back to the magnetosphere, an effect which will occur on the viscous timescale.

Secondly, if the photometric periods of CTTS are not directly due to the stellar rotation period, there is no reason to suppose that an individual photometric period will be compatible with the $v \sin i$ for that system. There are many cases in the literature for which photometric periods have been detected which, together with radius determinations from blackbody fitting, imply an equatorial velocity less than the $v \sin i$ found independently from the broadening of photospheric absorption lines. Since $\sin i$ cannot be greater than 1, this poses a problem for any explanation involving the stellar rotation period. In some cases the discrepancy is slight, and can be attributed to the large uncertainties associated with the radius determinations. In several cases though, the problem is quite severe and an alternative explanation is necessary. Furthermore, the bimodality seen in the distribution of photometric periods is not reflected in the $v \sin i$ data (Bouvier et al 1993).

An alternative explanation for this is that T Tauris rotate differentially with the poles where the spots form rotating slower than the equator (Smith 1994). This can only explain moderate differences in the $v \sin i$ and $P_{rot}$ as differ-ential rotation is not expected to be great (Johns 1995). If, however, the observed period is a beat frequency of the stars true rotational period and the Keplerian period at the magnetosphere then the discrepancy between period and $v \sin i$ is easily explained.

Thirdly, it is possible that CTTS could develop cool spots similar to those believed to exist on WTTS. These spots could then give rise to photometric periodicity, due to rotational modulation. If accretion is modulated at the same time at the beat frequency, two periods would be seen simultaneously. There are several cases of stars for which multiple periods have been detected. SU Aur was found by Herbst et al (1987) to have two periods of 1.55 and 2.73 days. These two periods, plus two more of 1.47 and 3.37 days were found by Bouvier et al (1993). Other examples include DR Tau, with periods of 2.8 and 7.3 days (Bouvier et al 1993), RY Lup [0.8 and 3.8 days, (Gahm et al 1989)] and AA Tau [2.6 and 4.2 days, (Vrba et al 1993) and at another time 8.2 days (Vrba et al 1993)].

## 4 OBSERVATIONAL CONSEQUENCES

How could the model suggested here be distinguished from the more conventional rotational modulation picture? Such a distinction is difficult to make, because the details of the magnetically funnelled accretion model which underlies both pictures are not certain. In particular, the exact origins of the emission lines are not well understood. Higher energy lines such as He I are presumed to originate close to the accretion shocks at the surface of the star (Guenther, personal communication). Lower excitation lines may well originate further out, either by reprocessing of the accretion luminosity, or by magnetic dissipation in the magnetosphere. However the lines originate, it is a reasonable assumption that lower energy lines will be formed further from the stellar surface than high energy lines. We can use this assumption to make some tentative predictions based on the geometry of the two pictures.

In the model we outline above, the low energy lines are expected to be formed at the outer edge of the magnetosphere. If magnetic dissipation excites the lines, then this effect occurs predominantly at azimuths where the interaction between the magnetic field and the disc is strongest. This means that the formation of the low energy lines is subject to the same beat frequency effects as the accretion rate. The variation in the lines should lead the continuum and high energy line variability by roughly the freefall time from the inner edge of the disc to the star's surface. This time will be several hours and so should be observable.

If magnetic dissipation is not important in forming the low energy lines, but these form instead by reprocessing of the accretion luminosity, then periodic variations of the low energy lines is still expected. The phase difference in this case will be approximately the light travel time from the star to the line emitting region, a matter of a few minutes.

In the case of rotational modulation, the behaviour of the lines is more difficult to predict, and has a greater dependence on the geometry and inclination angle of the system. If the photometric modulation is due to a hot accretion shock, then the magnetic and rotation axes would still have to be misaligned in order for the accretion luminosity to be modulated. This would mean that the magnetosphere boundary would be non-axisymmetric, and low energy lines formed by magnetic dissipation in the magnetosphere would still be formed preferentially at certain azimuths. In order for a periodicity to be observed in these lines, it is necessary for them to be eclipsed by passing behind the star. If the magnetosphere boundary lies several stellar radii from the star's surface, the system must be viewed close to edge on for this effect to occur. There should exist a number of systems for which the hot accretion shock is periodically eclipsed, while the low energy lines are not and do not vary with the same period.

If reprocessing is the dominant mode of low excitation energy line formation, then the situation is again different. The lines could be formed in the accretion column itself, or at the inner edge of the disc. In the latter case, a large area of the inner disc would be illuminated, and it is difficult to imagine any geometry and inclination angle which would produce periodic obscuration of the entire line forming region with the same period as the continuum. If the lines originate in the accretion column, it may be possible for periodic obscuration of the line forming region to occur, but this requires the inclination angle to be tightly constrained. Systems in which the low energy lines vary with the same period as the continuum would be expected to be in the minority. Any phase differences in this case would depend on the exact geometry of the accretion columns.

Spectroscopic studies of T Tauris have revealed that many lines undergo periodic variations with the same period as the photometry (Guenther 1994). Periodicity was also found in $H_\alpha$ for SU Aur by Johns et al (1992), although the period did not correspond to any of the photometric periods for this star. It is not yet clear how common such periodic line variability is, how the amplitude of the line variability changes with excitation energy, or whether there is a phase difference between the continuum and the lines.

## 5 CONCLUSIONS

The beat frequency modulated accretion model provides a new mechanism for explaining periodicities seen in the light curves of T Tauri stars which are accreting material from a circumstellar disk. It can therefore be applied to the CTTS, but not to the WTTS, which are thought not to actively accrete material from their disks.

The model provides an explanation for CTTS systems which display a period incompatible with their known $v \sin i$ values, as the beat period can be much longer than the rotation period of the star.

It also provides a framework in which to understand the behaviour of systems which change their photometric periods on timescales of years. Such behaviour is expected for a system accreting material from a region in its disc close to the corotation radius, as the beat period depends strongly on the exact orbit of the disc clump causing the modulated accretion.

It is possible that many of the periods detected for CTTS are in fact due to beat frequency modulated accretion, and therefore do not directly reflect the stellar rotation period. The cases in which two periods are seen simultaneously may then be instances in which the underlying stellar rotation is seen due to starspots, while the second period is due to modulated accretion at the beat period.

Finally, we note that if CTTS periodic behaviour is due to modulated accretion, rather than reflecting the stellar rotation period, then there is no reason to suppose that CTTS are spinning any slower than WTTS on average.

# 6  ACKNOWLEDGEMENTS

We thank Jim Pringle, Cathie Clarke, Eike Guenther, Matthew Bate and Derek Jones for helpful discussions. We are grateful to the anonymous referee for constructive criticism. We also thank the computer support staff at the Institute of Astronomy for maintaining the machines used in this work. KWS acknowledges a PPARC studentship.